\documentclass[12pt,letterpaper]{JHEP3}
\usepackage{epsf}

\newcommand{\littlefig}[2]{    \epsfxsize=#2cm    \epsfbox{#1}}

\title{De Sitter Holography and the Cosmic Microwave Background}

\author{Finn Larsen and Jan Pieter van der Schaar\\
        Michigan Center for Theoretical Physics\\
        Randall Laboratory, Department of Physics, University of Michigan\\
        Ann Arbor, MI 48109-1120, USA\\
        E-mail: \email{larsenf@umich.edu}, \email{jpschaar@umich.edu}}
\author{Robert G. Leigh\\
        Department of Physics\\
        University of Illinois at Urbana-Champaign\\
        Urbana, IL 61801\\
        E-mail: \email{rgleigh@uiuc.edu}}

\abstract{We interpret cosmological evolution holographically as a
renormalisation group flow in a dual Euclidean field theory, as
suggested by the conjectured dS/CFT correspondence. Inflation is
described by perturbing around the infra-red fixed point of the dual
field theory. The spectrum of the cosmic microwave background radiation
is determined in terms of scaling violations in the field theory. The
dark energy allows similar, albeit less predictive, considerations. We
discuss the cosmological fine-tuning problems from the holographic
perspective.}


\preprint{MCTP-02-09 \\ ILL-(TH)-02-02}

\begin{document}
\bibliographystyle{JHEP}

\section{Introduction}

Astronomical observations indicate that the universe is currently
accelerating. The simplest interpretation of this data is that we are
entering an epoch dominated by a positive cosmological constant, a de
Sitter phase, which will persist eternally. The Cosmic Microwave
Background anisotropy and other measurements pertaining to the very
early universe similarly suggest an epoch of rapid acceleration in our
distant past. The history of the universe may thus be interpreted as the
interpolation between two de Sitter phases. The traditional framework to
interpret this evolution is to hypothesize some medium, the inflaton
field or quintessence, and design the dynamics of the corresponding
scalar field so it dominates at early and late times, respectively.
Recently, a more radical interpretation was proposed: the cosmic
evolution is {\it holographic}; {\it i.e.} it can be usefully analyzed
in terms of a dual field theory \cite{Strominger-inflation, BBM-dSmass}. 
This paper develops the holographic proposal further.

A holographic duality has been established only for anti-de Sitter space
(AdS) \cite{AGMOO-AdS}, but it is plausible that many other backgrounds
similarly allow holographic representations, particularly in view of the
area law for gravitational entropy. In the case of AdS the dual theory
is a conformally invariant gauge theory, situated at the AdS boundary at
large radius. The symmetries of de Sitter space (dS) suggest that their
duals are conformal field theories (CFT's) as well; these theories are
Euclidean and located at asymptotically early (or late) times
\cite{Hull-dS/CFT, Strominger-dS/CFT}. 
If the dS/CFT correspondence is correct, the
cosmological evolution interpolating between two de Sitter-phases can be
naturally interpreted as the renormalization group flow linking two
CFT's (analogous flows were considered in AdS
\cite{BVV-holrg,FGPW-rgf,PS-AdSrgf,ST-AdSrgf}). Then there is one
underlying quantum field theory, with the future de Sitter phase
governed by the ultraviolet (UV) fixed point, and the past corresponding
to the infrared (IR) fixed point. The increasing entropy of the
expanding universe is interpreted holographically as the ``integrating
in'' of the reverse renormalization group flow.

The purpose of this paper is to make this abstract scenario more
concrete. Our main example is the anisotropy of the cosmic microwave
background (CMB) radiation. We find that holography gives the same
predictions as standard slow-roll inflation; in fact, there is a simple
dictionary between the holographic model and the standard approach using
a single spacetime scalar field. However, the holographic interpretation is
different: the basic input is the dual quantum field theory, rather than
the inflationary potential; so now scaling violations are interpreted in
terms of the $\beta$-function and the anomalous dimension. One reason
this is interesting is that it gives a new perspective on the
fine-tuning problems introduced by inflation, and on the cosmological
constant problem. A different approach to interpreting the CMB in terms of 
critical scaling behavior can be found in \cite{AMM-CFT/CMB}.

Another motivation for studying the holographic formulation is that it
may provide a phenomenologically viable path to 
microphysics\footnote{Other recent attempts to connect cosmology and 
microphysics appear in \cite{ADD-cosmo, BDL-bwcosmo, Copeland, KKLS-cmb, 
Ekpyrotic}.}. 
It is often presumed that the inflationary potential will be the interlocutor,
inferred in some detail from observations as well as derived from a low
energy limit of string/M-theory. It seems to us no less fanciful to
envision interpreting observations in terms of scaling violations and
compare them with a microscopic understanding of the dual field theory.

In the simplest viable models of inflation, there is a single inflaton
field which dominates the energy density \cite{AS-newinf, Linde-newinf}.
The quantum fluctuations of this field are ultimately responsible for
the CMB anisotropy. The detailed spectrum depends on small departures
from pure de Sitter space and is conventionally parametrized in terms of
the slow-roll parameters $\epsilon$ and $\eta$. In the dual
interpretation the inflationary epoch is reinterpreted as a quantum
field theory close to its IR fixed point. The CMB spectrum now depends
on the scaling violations of the field theory and is simply parametrized
by the $\beta$-function and the anomalous dimension $\lambda$. 
Perturbative quantum field theory gives $\beta\ll\lambda$
corresponding to the prediction $\epsilon\ll\eta$.

In Section 2, we first review standard inflationary cosmology and then
summarize the conjectured dS/CFT correspondence. Next, we explain how
the bulk physics determines properties of the CMB, and show how this is
reinterpreted in the dual theory. Of course, we do not as yet understand
all aspects of the dS/CFT duality, but for our purposes a few simple
features suffice. The most conservative view is that we simply
reorganize the standard inflationary picture in a way inspired by dS/CFT,
but ultimately interpretable in terms of symmetries. This point of view
could be useful phenomenologically even if no holographic dual exists.
In the remainder of the paper we will simply assume the dS/CFT
correspondence and take the view that we are deriving some of its
consequences.

In Section 3 we introduce a concrete quantum field theory realization of
the central features assumed in our general discussion. We explicitly
compute the CMB spectrum in this toy model of the early universe. We
conclude the paper in Section 4 with a discussion of some standard
problems in inflationary cosmology, such as the cosmological constant
problem, the fine-tuning of the inflaton potential, and the initial
value problem. We translate these problems to the dual theory and
explain some potential advantages of the holographic perspective.

\section{Holographic Cosmology}

\subsection{FRW Cosmology}

We consider 4d Einstein gravity coupled to scalars $\phi^I$ {\it via}
\begin{equation}
{\cal L}={1\over 2 \kappa^2} R - {1\over 2} G_{IJ} \partial_\mu \phi^I
\, \partial^\mu \phi^J - V(\phi^I) ~,
\label{lagrangian}
\end{equation}
where $\kappa^2= 8\pi G= 1/{M_p}^2$ and $G_{IJ}$ is the metric on the
moduli space of the scalars $\phi^I$. The potential $V(\phi^I)$
is assumed to have de Sitter extrema. Using the standard spatially
flat FRW metric 
\begin{equation}
ds^2= -dt^2 + a(t)^2 \left( dr^2 + r^2 d\Omega_2^2 \right)~,
\label{FRWansatz}
\end{equation}
and assuming spatial isotropy $\vec{\nabla} \phi^I=0$, the equations
of 
motion are given by
\begin{eqnarray}
\ddot{\phi}^I &+& 3 H \dot{\phi}^I + G^{IJ} {dV \over d\phi^J} 
= 0 ~, \label{scalareom} \\
H^2 &=& {1\over 3} \kappa^2 \rho ~, \label{H2eom} \\
\dot{H} &=& -{1\over 2} \kappa^2 \left( \rho + p \right) ~,
\label{Hdoteom}
\end{eqnarray}
where the Hubble parameter $H$ is $H\equiv \dot a/a$, the dot
representing a derivative with respect to the coordinate time $t$, and
the density $\rho$ and pressure $p$ are given by
\begin{eqnarray}
\rho &=& {1\over 2} G_{IJ} \dot{\phi}^I \dot{\phi}^J + V(\phi^I)~,
\nonumber \\
p &=& {1\over 2} G_{IJ} \dot{\phi}^I \dot{\phi}^J - V(\phi^I)~.
\label{rhoand}
\end{eqnarray} 
The equations of motion give
\begin{equation} 
{3\over 4} \kappa^2 (\omega+1) = H^{-2}G^{IJ} {d H \over 
d\phi^{I}} {dH \over d\phi^{J}} ~, \label{eos}
\end{equation}
for the equation of state parameter $\omega \equiv p / \rho$. Also
recall that one of the equations of motion is redundant, {\it e.g.}
(\ref{scalareom}) and (\ref{H2eom}) imply (\ref{Hdoteom}).

In the following we drop the index $I$ and consider a single scalar with
canonical kinetic energy; the generalization back to multiple scalars is
straightforward.

\subsection{The dS/CFT Correspondence}

We now want to reinterpret this 4d bulk cosmology in terms of a dual 3d
Euclidean quantum field theory. The idea is that the scalar field breaks
the de Sitter symmetry in the bulk and this will correspond to broken
scale invariance in the dual theory.

According to the dS/CFT correspondence a scalar field $\phi$ with
asymptotic value $\phi_0(\vec x)$ is dual to an operator ${\cal O}$.
This means the dual theory is perturbed away from its conformal fixed
point
\begin{equation}
{\cal L}= {\cal L}_{\rm CFT} + g {\cal O}~,    
\end{equation}   
where the coupling $g=\kappa\, \phi_{0}$. The QFT scale parameter is
$\mu \propto a$, mapping bulk UV (early times) to field theory IR and
{\it vice versa}. With this map the QFT $\beta$-function translates into
the bulk as
\begin{equation}
\beta \equiv {\partial g \over \partial \log \mu}=
{\partial \over \partial \log a} \kappa \phi =
-{2 \over \kappa H}  {d H \over {d\phi}}~. \label{betafn}
\end{equation}
The last step employed the FRW equations (\ref{H2eom}-\ref{Hdoteom}).

To proceed we assume that the potential has a mass term
$\partial^{2}_{\phi}V|_{\phi=0}=m^{2}\neq 0$. Then the Klein-Gordon
scalar equation (\ref{scalareom}) with $a(t)\sim e^{H_{0}t}$ as either
$t\to\infty$ or $t\to -\infty$ determines the asymptotic wave function
$\phi =\phi_{0}(\vec{x}) e^{\lambda H_0 t}$ where
\begin{equation}
\lambda^2 + 3 \lambda + {m^2 \over H_0^2} = 0 ~,
\label{lambdam}
\end{equation}
giving two solutions for $\lambda$
\begin{equation}
\lambda_{\pm} = -{3\over2} \pm \sqrt{\left({3\over2}\right)^2 -
{m^2 \over H_0^2}} ~.
\label{lambdapm}
\end{equation}
A similar relation (with $m^{2}\to -m^{2}$) is standard in the AdS/CFT
correspondence \cite{Witten-AdS}. To interpret the parameter $\lambda$
note that, in the asymptotic de Sitter space, the wave function
transforms as $\phi \to \phi e^{\lambda H_0 \Delta t}$ under a time
translation $t \to t+\Delta t$. But, according to the metric
(\ref{FRWansatz}), this can also be viewed as a rescaling of the spatial
slice $\vec{x}\to ({\mu\over\mu_{0}})\vec{x}$ where
${\mu\over\mu_{0}}=e^{\Delta t H_{0}}$. Thus $\phi_{0}\to
({\mu\over\mu_{0}})^{\lambda}\phi_{0}$ under rescaling within the
spatial slice; so $\lambda$ is the anomalous scaling dimension of the
dual operator ${\cal O}$. We can relate it to the beta function as
\begin{equation}
\lambda = {\partial \log \phi_{0}\over \partial \log a} =
{\partial \log g\over \partial \log\mu} = {\beta\over g}~.
\label{deflambda}
\end{equation}

For each field there are two values of $\lambda$, given by
(\ref{lambdapm}). The interpretation of $\lambda$ in terms of the
asymptotic behavior of the wave function $\phi_{\pm}\sim
e^{\lambda_{\pm} H_0 t}$ shows that $\lambda_{+}$ corresponds to
infinite asymptotic energy 
$E=\int \dot{\phi}^{2}a^{3} d^{3}x=\infty$; in contrast,
$\lambda_{-}$ is a finite energy perturbation $E<\infty$. It is
therefore natural to interpret the $\lambda_{+}$ as a deformation of the
dual field theory whereas $\lambda_{-}$ corresponds to an excitation of
the theory without changing the underlying Lagrangian. This
interpretation is standard in the AdS/CFT correspondence \cite{BKL-AdS}.
Since the energetics is similar in de Sitter space we expect the result
to hold here as well\footnote{ The distinction usually made is actually
between {\it normalizable} and {\it non-normalizable} states. In bulk
the standard norm is the Klein-Gordon norm, but in de Sitter space this
norm vanishes because the wavefunctions $\phi_{\pm} \sim
e^{\lambda_{\pm} H_0 t}$ are real. Using the asymptotic energy as
criterion attempts to circumvent this difficulty.}. In our application
we want to consider the RG-flow between different theories. This will
require deforming the theory and thus considering $\lambda_{+}$.

Let us now focus on the IR theory at $t\to -\infty$, the traditional
inflationary regime. The approach to an IR fixed point is described by
an {\it irrelevant} deformation of the IR theory. The field theory
expectation is thus for a positive anomalous dimension
$\lambda^{IR}_{+}>0$. According to (\ref{lambdapm}) this corresponds to
$m^{2}<0$, a tachyon mass. This is expected from the bulk point of view
as well: the scalar field rolls down a hill from a maximum.

The approach to the IR fixed point of a quantum field theory is
generally described by many irrelevant operators. However, the final
approach is dominated by the operator(s) of smallest dimension because
this will have survived the RG flow the longest. If this operator is
nearly marginal, with $\lambda^{IR}_{+}\ll 1$, then the slow-roll
conditions for inflation are satisfied in the bulk, as we will show in
the next subsection.

The assumption of a nonvanishing mass of the bulk scalar is significant
for the preceding discussion. The corresponding field theory statement
is $\beta\propto g$ for small coupling, as follows from
(\ref{deflambda}) with constant $\lambda$. This behavior corresponds to
the $\beta$-function of a coupling with classical mass dimension, a
common behavior in three dimensions. If we want to consider instead
couplings with $\beta\propto g^{n}$ for $n>1$ the corresponding bulk
field has $m^{2}=0$ and the anomalous dimension $\lambda_{+}$ of the
dual operator vanishes at the fixed point. In this case (\ref{betafn})
can be inverted to determine the spacetime potential as $V\propto
\phi^{n+1}$ for small $\phi$. In perturbative QFT $\lambda$ always
satisfies (\ref{deflambda}); so we {\it define} $\lambda$ as the $g$
dependent function
\begin{equation}
\lambda \equiv {\beta\over g}~,
\label{lambdadef}
\end{equation}
away from the fixed point. Then our considerations will be valid 
for these more general theories as well.

To complete our dictionary we introduce the holographic $c$-function.
This function decreases along the RG-flow towards the IR and thus
provides a precise formulation of the heuristic idea that the RG-flow
integrates out degrees of freedom. The $c$-function has not yet been
firmly established in QFT above two dimensions but the evidence in its
favor is substantial \cite{AFGJ-ctheorem}. Importantly, there is a natural
candidate for a holographic interpretation in AdS \cite{FGPW-rgf} as well as in
de Sitter \cite{Strominger-inflation}; it is given in 4 dimensions by
\begin{equation}
c\equiv {1 \over \kappa^2 H^2}~. \label{holc}
\end{equation}
The FRW equation (\ref{Hdoteom}) gives $\dot{H}<0$ for matter satisfying
the null energy condition ({\it e.g.} $p+\rho>0$) so $c(t)$ increases in
time. In cosmology it expresses the increasing entropy of the
expanding universe; in QFT it represents the ``integrating in'' along
the RG-flow towards the UV fixed point. De Sitter RG-flows and 
the holographic ``c-theorem'' were recently discussed 
in \cite{LMM-talldS}.

\subsection{The Slow-Roll Parameters}

Spacetime inflates when the cosmological evolution is dominated by the
potential energy of the scalar (for reviews on inflation see
\cite{Liddle-Lyth, Carroll-TASI, Steinhardt, Lyth-Riotto}) . This
situation arises when $\dot \phi^2 \ll V(\phi)$ and $|\ddot \phi| \ll
|3H\dot \phi| , |\partial_{\phi}V|$. It is convenient to introduce the
slow roll parameters\footnote{Our slow-roll parameters are denoted
$\epsilon_H$ and $\eta_H$ in [22]. Another set of commonly used
parameters are related to ours as $\epsilon_{\rm here} = \epsilon_{\rm
there}$ and $\eta_{\rm here} = \eta_{\rm there}-\epsilon_{\rm there}$ in
the slow-roll limit.} \cite{Liddle-Lyth}
\begin{eqnarray}
\epsilon &\equiv& -{\partial \log H \over \partial \log a} = {2
\over 
\kappa^2} \left( {1\over H}{dH \over d\phi}  \right)^2~,
\label{epsilon} \\
\eta &\equiv& -{{\partial \log {\partial H \over \partial \phi}} \over
\partial \log a} = {2 \over \kappa^2} {1 \over H} {d^2 H
\over d\phi^2}~, \label{eta}
\end{eqnarray}
and write these conditions as
\begin{equation}
|\epsilon|,|\eta| \ll 1 ~.
\label{slowroll}
\end{equation}
Inflation ends precisely when the slow-roll conditions are violated.
From then on the cosmological evolution will be dominated by the kinetic
energy of the scalar and by matter and radiation.

We now want to express the slow-roll parameters in terms of the
$\beta$-function (\ref{betafn}) and the anomalous dimension $\lambda$
(\ref{lambdadef}), the natural parameters in the dual field theory.
Comparing (\ref{betafn}) and (\ref{epsilon}) gives
\begin{equation}
\epsilon = {1\over 2} \beta^2 ~.
\label{epsilon-beta}
\end{equation}
The other slow-roll parameter $\eta$ (\ref{eta}) can be written as
\begin{equation}
\eta = {1\over 2}\beta^2 -\frac{1}{\kappa}{d\beta
\over d\phi} ~.
\label{etandepsilon}
\end{equation}
In the vicinity of the fixed point at $g=0$  (\ref{deflambda}) then gives
\begin{equation}
\eta = - \lambda ~.
\label{approxeta}
\end{equation}
When $\lambda$ vanishes close to a fixed point with $\beta\propto g^{n}$
for $n > 1$ the relation becomes $\eta = - n \lambda$ to the leading order.
The relations (\ref{epsilon-beta}) and (\ref{approxeta}) provide the
basic dictionary between scaling violations in QFT and inflation.

Inflation ends when the slow-roll parameters approach unity. The dual
field theory interpretation is that the theory enters a complicated
interacting regime with no useful perturbative description. A plausible
scenario is that when $\lambda\ll 1$ is violated many hitherto
negligible operators become important and so the simple scaling regime
breaks down. The end of inflation can be modeled in detail but one must
make specific assumptions about the complete field content and the
corresponding $\beta$'s and $\lambda$'s, even away from the weakly
coupled regime. We expect that such models would parallel the standard
inflationary models.

\subsection{Holography and the CMB}

The main application of these ideas is to cosmological perturbations. A
massless scalar field in de Sitter space has fluctuations at each
wavenumber $k$, with magnitude $\delta\phi_k ={H \over 2\pi}$. This
leads to density perturbations on every scale if the potential is
sufficiently flat. The power spectrum of the scalar component of the
spatial metric is
\begin{equation}
{\cal P}_{\rm scalar} = \left({H\over \dot{\phi}}\right)^{2} 
\left. \left({H\over 2\pi}\right)^{2}\right|_{k=aH} \propto k^{n_S -1} ~, 
\label{scalarfluc}
\end{equation}
where the subscript $k=aH$ indicates that the expression is to be
evaluated at the moment when the physical scale of the perturbation is
equal to the Hubble radius. Perfect scale invariance corresponds to
$n_S=1$; and the scaling violations, due to the slow rolling of $H$ and
$\dot{\phi}$, become
\begin{equation}
n_S = 1-4\epsilon + 2\eta ~.
\label{scalarin}
\end{equation}
The tensor component of the spatial metric also develops fluctuations,
interpreted as graviton waves. Their power spectrum is
\begin{equation}
{\cal P}_{\rm grav} =
2\kappa^{2} \left. \left({H\over 2\pi}\right)^{2} \right|_{k=aH} 
\propto k^{n_T} .
\label{tensorfluc}
\end{equation}
Perfect scale invariance now corresponds to $n_T=0$; the scaling 
violations are 
\begin{equation}
n_T= -2\epsilon ~.
\label{tensorin}
\end{equation}
A third observable is the ratio $r$ of the tensor and scalar
power\footnote{The ratio of multipole moments $C_{l}^{(T)}/C_{l}^{(S)}$
is often quoted instead. It differs from this by an $l$-dependent
numerical factor of geometric origin.}
\begin{equation}
r \equiv {{\cal P}_{{\rm grav}}\over {\cal P}_{\rm scalar}} = 
2\kappa^2 \left( { \dot{\phi} \over H }
\right)^2 = 4\epsilon ~.
\label{TSratio}
\end{equation}

The observables (\ref{scalarfluc}),(\ref{tensorfluc}),(\ref{TSratio})
are all expressed in terms of the slow-roll parameters $\epsilon$ and
$\eta$; they can therefore also be expressed in terms of the dual
$\beta, \lambda$, using (\ref{epsilon-beta}) and (\ref{approxeta}). The
results to the leading significant order are
\begin{eqnarray}
n_S &=& 1- 2\lambda ~,\\
n_T &=& -\beta^{2}~,\\
r &=& 2\beta^{2}~,
\end{eqnarray}
with $\beta\ll\lambda\ll 1$. The power of the tensor component is thus
negligible; and the tilt of the scalar component is determined in terms
of $\eta = -\lambda$ alone and because $\lambda > 0$ this implies a
slightly red spectrum. In the standard notation $\epsilon\ll\eta\ll 1$
and $\eta < 0$, results also predicted by a large class of inflationary
models \cite{Liddle-Lyth, Steinhardt, Lyth-Riotto}.

We would like to understand these results from the QFT. The first step
is to justify the use of perturbation theory. In QFT inflation ends in a
strongly coupled regime; we want to compute correlators at this time
with arguments corresponding to astronomically large distances. A huge
RG transformation of ${\cal O}(10^{50-60})$ transforms this into
correlators at more natural distances, and the RG flow also shifts the
strong coupling towards the IR by a large amount, presumably bringing
the coupling to the perturbative regime\footnote{This might not happen,
despite the large scale transformation. Inflationary models with
predictions different from ours might arise this way.}. (However, we are
not requiring the fixed point theory itself to be weakly coupled.)

The next step is to derive the fluctuation spectrum directly from QFT.
This should also be possible if we assume certain details of the dS/CFT
duality to hold. This is apparent, since the bulk fluctuations are
determined by properties of correlation functions. The asymptotics of a
bulk correlation function $\langle \phi\phi\rangle$ will be related to a
boundary correlation function $\langle{\cal O}{\cal O}\rangle$, if
$\phi$ is dual to ${\cal O}$. The momentum dependence of the  boundary
correlation function, close to the fixed point, is determined by the
scaling dimension of ${\cal O}$ at the fixed point. Similar computations
for the energy-momentum tensor in the CFT correspond to tensor
fluctuations in bulk. We have not carried out this procedure explicitly;
it would clearly be interesting to do so.

\subsection{Dark Energy} Until this point our focus has been on the
approach to the IR fixed point, {\it i.e.} the inflationary regime. In
holographic cosmology the entire cosmological evolution can be
interpreted in the dual field theory, at least in principle. The
holographic variables are not useful throughout the RG-flow but they
might simplify also close to the UV fixed point, {\it i.e.} in the
regime dominated by dark energy.

The UV theory is perturbed by a relevant operator, inducing the RG flow.
The simplest is to assume that one such operator dominates. This means
there is some scalar field $\phi$ satisfying
\begin{equation}
\phi \sim e^{\lambda^{UV}_{+} H_0 t}~~~~;~~t\to\infty ~,
\label{plusUV}
\end{equation}
with\footnote{We only consider real $\lambda^{UV}_{+}$.} $-{3\over
2}<\lambda^{UV}_{+}<0$. Equivalently, the mass of the UV scalar field is
expected in the range $0<{m^{2}\over H_0^2}< ({3\over 2})^{2}$.

This type of model would be very similar to the quintessence model of
dark energy \cite{RP-scalarcos, WCOS-quint}. As for inflation one can
translate the natural holographic variables, $\beta$ and $\lambda$, to
those customary in the quintessence literature. There one usually
concentrates on the equation of state parameter $\omega$, as well as the
scalar mass $m$. $\lambda$ is related to $m^{2}$ through
(\ref{lambdapm}) and $\omega$ is given by $3(\omega+1)=\beta^2$ . We can
now take over standard considerations from quintessence.

An important difference between inflation and dark energy is that the
holographic perspective gives no clear motivation for considering a
theory of a single scalar field close to the UV limit. Indeed, although
we have used the same notation $\phi$ for the scalar field at both $t\to
-\infty$ and $t\to\infty$ they are generally completely unrelated; the
details of the UV and IR theories are very different. The only relation
between these two regimes is that, as a matter of principle, they can be
described by the same underlying quantum field theory.

\section{The Dual Field Theory}

To this point we have not discussed in detail the precise nature of the
duality between de Sitter space and conformal field theory
\cite{Strominger-dS/CFT,SSV-review, Klemm}. Of course it could be a
full-fledged duality, similar in nature to AdS/CFT. This view has
conceptual \cite{Fischler, Susskind, Witten-dS} and practical difficulties: 
for example, de
Sitter space does not seem to have simple realizations in string theory
\cite{BHM-dSIIB, HHK-dScos,KLPS-dScos,Kallosh-dSsusy, PetkouSiopsis, 
Silverstein-dS, Townsend-MQ}.
It has been established that the dS/CFT correspondence, if true, is not
simply a continuation of the AdS/CFT correspondence \cite{BMS-dS/CFT, 
SV-dS/CFT}.

On the other hand, it is sometimes claimed that most or all gauge
theories have gravitational duals \cite{Polyakov}; this could be 
taken to mean that any
Euclidean CFT will have a dual de Sitter description. The matter content
and other properties of the spacetime theory will then depend on the
choice of CFT, but the background geometry should be de Sitter for all
such CFT's. This optimistic view motivates us to consider a toy model,
the simplest example of a Euclidean theory in 3 dimensions with
properties of the kind we have assumed in this paper.  That the central
charge $c_{IR}\sim \kappa^{-2}\Lambda_{IR}^{-1}$ is rather low for
typical $\Lambda_{IR}$ indeed gives evidence that the IR CFT can be
quite simple. Even if this toy model is in fact not dual to de Sitter
space it is an instructive illustration of our ideas.

Let us then consider the model
\begin{equation}\label{eq:ircftS}
S_E=\int d^3x\ \left[ \frac12 (\partial_\mu\varphi)^2 +
\frac{g}{6!}\varphi^6\right]~.
\end{equation}
This has the virtue of being renormalizable. The $\varphi^6$ term is
classically marginal but at the quantum level the coupling $g$ is in
fact marginally irrelevant in the IR. This is just what we want: the IR
CFT is the trivial fixed point and scaling violations away from the
fixed point are governed by perturbation theory.

The renormalization properties of this theory are quite simple in
perturbation theory. Of course, for the action (\ref{eq:ircftS}) to be
valid near the fixed point, it must be finely tuned, as there are
several relevant perturbations. (This feature could
presumably be fixed in a simple supersymmetric extension of the model.)
The anomalous dimension of $\varphi^6$ can be computed by considering
the correlation function $\langle T_{\mu\nu}(x)\varphi^6(y)\rangle$ or,
equivalently by considering $G_{66}(x,y)=\langle
\varphi^6(x)\varphi^6(y)\rangle$, which satisfies the Callan-Symanzik
equation
\begin{equation}
\left[ 
\frac{\partial}{\partial\ln\mu}+\beta_g\frac{\partial}{\partial
g}-2\lambda_{(\varphi^6)}\right] G_{66}=0~.
\label{g66int}
\end{equation}
At leading order the diagram

\centerline{\littlefig{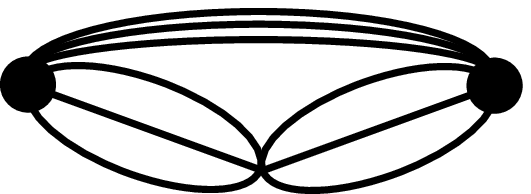}{4}}
\smallskip

\noindent gives
\begin{equation}
G_{66}\sim 1+\frac{20}{3!}g\int
\frac{d^3k}{(2\pi)^3}\frac{d^3p}{(2\pi)^3}\
\frac{1}{k^2}\frac{1}{p^2}\frac{1}{(p+k)^2}~,
\label{g66exp}
\end{equation}
and thus
\begin{equation}
\lambda_{(\varphi^6)}=\frac{5}{3}\frac{g}{16\pi^2}~.
\label{toylam}
\end{equation}
The general relation (\ref{lambdadef}) gives the $\beta$-function
\begin{equation}
\beta_g=\frac{5}{3}\frac{g^2}{16\pi^2}~.
\label{toybeta}
\end{equation}
This result also follows from explicit computation of the diagram 

\smallskip
\centerline{\littlefig{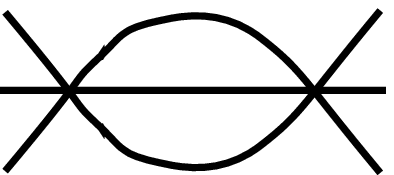}{2.5}}
\smallskip

\noindent which leads to the same integral as (\ref{g66exp}).

Inserting the results (\ref{toybeta}) and (\ref{toylam}) for $\beta$ and
$\lambda$ into (\ref{epsilon-beta}) and (\ref{approxeta}) we find
expressions for $\epsilon$ and $\eta$ and hence for the CMB spectrum. We
find at weak coupling $g\ll 1$ that $\epsilon\ll\eta$, in agreement with
the general discussion in Section 2.4. Of course in the explicit model
the conformal dimension is derived in terms of the coupling $g$, a
microscopic parameter.

We stress that these results should not be taken as a prediction of
holography; they are the predictions derived from this specific dual
theory which we have motivated only by its simplicity. The explicit
discussion is meant to demonstrate how holographic considerations can
lead to a concrete computational framework.

The theory (\ref{eq:ircftS}) is not completely satisfying as an IR
theory for several reasons. We have already mentioned that the existence
of relevant perturbations lead to a fine tuning: this can be interpreted
as a very particular renormalization group trajectory for which the
action is valid. We will comment further on this issue in the following
section. Another potential problem with this IR CFT is that it is weakly
coupled: we of course chose this to be true to facilitate explicit
computations, but it is presumably not expected to be the case, if the
CFT is to be dual to a weakly coupled gravitational system.

\section{Discussion}

One of the motivations for considering holographic cosmology is that it
offers an alternative viewpoint on several cosmological fine-tuning
problems and may ultimately lead to their resolution. As a conclusion to
the paper we discuss in turn the fine-tuning of the inflationary
potential, the cosmological constant problem, and the initial value
problem.

{\bf Fine-tuning of the Inflationary Potential:} In the usual
implementation of inflation, phenomenological constraints force the
inflationary potential to be extremely flat. For example, in a model
with $V(\varphi) = - {1\over 4}g \varphi^{4}$, one must have $g\sim
10^{-13}$ \cite{Lyth-Riotto}. Such potentials are regarded as unnatural
in quantum field theory: dimensionless coefficients are expected to be
roughly of ${\cal O}(1)$. If one tries to choose them much smaller, they
are expected to take their natural values after renormalization.

The holographic perspective improves this situation. Since we interpret
the inflationary epoch as the final approach to the IR fixed point after
a long flow from a complicated UV theory, there are naturally few
degrees of freedom. Any fine tuning of the bulk theory has an
interpretation in terms of the properties of this IR flow. A viable IR
fixed point must possess suitable marginally irrelevant perturbations,
and the RG flows must come in to the fixed point along these directions.
The absence of relevant perturbations at the fixed point may be a
desirable  property from this  point of view, although it is not clear
if this is realistic. Modifications to the bulk inflaton potential do
not correspond to quantum effects in the dual theory, but rather to
changes in the properties of the fixed point.  A nice  property is that,
in a given theory, universality of RG flows guarantees that the details
of the UV region are not too important for inflation, apart from the
fact that in a given model, particular trajectories may be  required.
This robustness of the IR theory thus appears to constitute a resolution
of this fine-tuning problem.

{\bf The Cosmological Constant Problem:} In the holographic model the
cosmological evolution is an inverse RG flow from small $c_{IR}$ (large
cosmological constant) in the past to large $c_{UV}$ (small cosmological
constant) in the future. The ratio
$$
{c_{UV}\over c_{IR}} = {\Lambda_{IR}\over\Lambda_{UV}} \sim
10^{120}~,
$$
is enormous and the holographic perspective does not require this to be
so (there are perfectly respectable RG flows with $c_{UV}/c_{IR}\sim
{\rm a~few}$). Holographic cosmology therefore does not resolve the
cosmological constant problem.

On the other hand, a possible motivation for large $c_{UV}/c_{IR}$ may
be the following. In the AdS/CFT correspondence classical space-times
correspond to the large $N$ limit, and thus to a large value of the
central charge. In the holographic scenario, the value of the
cosmological constant is rather directly related to the complexity of
the Universe, or more precisely, the complexity of the dual field
theory. Large $N$ in the AdS/CFT example may be a simple prototype for
complexity in the Universe. There is thus a clear anthropic account of
the cosmological constant here: only those Universes with sufficient
complexity are realistic, and this corresponds to a large dual central
charge, and a small cosmological constant.

A technical aspect of the usual cosmological constant problem is that
small values are unstable to quantum corrections. It is important to
contemplate the role of quantum corrections in the holographic scenario.
Given a suitable starting point in the UV of the dual field theory,
quantum effects simply generate the RG flow. The details of how large
these corrections are do not matter in detail: the natural value for
$\Lambda$ is approached in the IR. One could also consider modifications
to the UV theory, but as long as $c_{UV}$ is large, they do not
de-stabilize $\Lambda$. Thus, as long as the whole scenario exists, {\it
the technical cosmological constant problem is resolved}.
For other holographic discussions of the cosmological
constant problem see \cite{Banks-Lambda, Bousso-Lambda, VV-rgcos}.

{\bf The Initial Value Problem:}
Inflationary models usually take the view that inflation was a finite
epoch. Singularity theorems state that the spacetime preceding this
epoch inevitably has a singularity somewhere \cite{BGV}, presumably to be
resolved by string theory/M-theory. In principle, the pre-inflationary
epoch provides the initial conditions for inflation; but in practice
this data is inflated away, allowing inflation to make predictions that
are independent of the initial conditions. The initial value problem is
therefore not acute, but the situation is nevertheless conceptually
unsatisfactory.

In holography the situation is quite different. In this paper, we have
effectively assumed that the pre-inflationary period is eternal
de Sitter, as the endpoint of the dual RG flow was assumed to be the IR
fixed point. Technically, the singularity theorems do not apply to
eternal de Sitter but it is far from clear that de Sitter constitutes a
satisfactory boundary condition.  
One problem is that we must somehow
determine amplitudes for all irrelevant operators in the theory. Another
(presumably related) issue is that, after spatially inhomogeneous field
configurations are taken into account, de Sitter may be unstable towards
fragmentation \cite{Linde, Starobinsky}. 

A variation of our scenario is to identify the inflationary epoch in the
dual theory as a fortuitously close approach (with sufficient
e-foldings) to a fixed point with relevant directions. If this were the
case, then the pre-inflationary epoch would correspond to some further
RG flow; and the properties of this flow would constitute everything
there is to the initial value problem.

Of course, in either formulation, the initial values are observationally
irrelevant. As we have shown, a single operator typically dominates, and
we can make predictions without knowing its detailed properties. Thus
the practical situation is completely parallel to standard inflation.

\acknowledgments
We thank S. Carroll, K. Freese, L. Kofman, E. Poppitz, A. Peet, 
M. Spradlin, and R. Wald for discussions. RGL thanks the University of 
Michigan for hospitality and FL thanks the theory group at the University 
of Chicago for their comments and for hospitality. Supported in part by 
the U.S. Department of Energy under grants DE-FG02-91ER40677 and 
DE-FG02-95ER40899.

\bibliography{holcmb}

\end{document}